\documentclass[aps,pre,twocolumn,showpacs,amsmath,graphicx]{revtex4}

\usepackage{array}
\usepackage{comment}
\usepackage{subfigure}
\usepackage{graphicx}
\usepackage{multirow}
\usepackage{subfigure}
\newcommand{\rd}{random}
\newcommand{\rc}{random cascading blocks}
\newcommand{\cb}{cascading dead blocks}
\newcommand{\bd}{dead blocks}

\begin{document}
\preprint{This line only printed with preprint option}
\title{Scaling of Earthquake Models with Inhomogeneous Stress Dissipation}

\author{Rachele Dominguez}
\email{erg.dominguez@wku.edu}
\affiliation{Department of Physics and Astronomy, Western Kentucky University, Bowling Green, Kentucky 42101, USA}

\author{Kristy Tiampo}
\affiliation{Department of Earth Sciences, University of Western Ontario, London, Ontario, N6A 5B7 Canada}

\author{C. A. Serino}
\affiliation{Department of Physics,  Boston
University, Boston, Massachusetts 02215, USA}

\author{W. Klein}
\affiliation{Department of Physics and Center for Computational Science,  Boston
University, Boston, Massachusetts 02215, USA}

\begin{abstract}
Natural earthquake fault systems are highly non-homogeneous. The
inhomogeneities occur because the earth is made of a variety of materials
which hold and dissipate stress differently. In this work, we study scaling in earthquake fault models which 
are variations of the Olami-Feder-Christensen (OFC) and Rundle-Jackson-Brown (RJB) models. We use the scaling to explore the effect of spatial inhomogeneities due to damage and inhomogeneous 
stress dissipation in the earthquake-fault-like systems when the stress transfer 
range is long, but not necessarily longer than the length scale
associated with the inhomogeneities of the system. We find that the scaling depends not only on the amount of damage, but also on the spatial distribution of that damage. \end{abstract}
\maketitle

\section{introduction\label{sec:introduction}}
The spatial arrangement of fault inhomogeneities is dependent on the geologic history of the fault. Because this
history is typically quite complex, the spatial distribution of the
various inhomogeneities occurs on many length scales. One way that
the inhomogeneous nature of fault systems manifests itself is in the
spatial patterns which emerge in seismicity graphs~\cite{tiampo_mean-field_2002,tiampo_ergodicity_2007}.

Despite their inhomogeneous nature, real faults are often modeled
as spatially homogeneous systems. One argument for this approach is
that earthquake faults have long range stress transfer~\cite{klein_structure_2007}, and if this range
is longer than the length scales associated with the inhomogeneities
of the system, the dynamics of the system may be unaffected by the
inhomogeneities. However, it is not clear that this is the case. Consequently it is important to investigate the situation in which the stress transfer range is comparable to or less than the length scales associated with the damage or stress dissipation inhomogeneities. 

In
this work, we study scaling in cellular automaton models of earthquake faults. We use a variation of a model introduced 
initially by Rundle, Jackson and Brown (RJB) and re-introduced independently by Olami, Feder and Christensen
(OFC) to explore the effect of
spatial inhomogeneities in earthquake-fault-like systems when stress transfer 
ranges are long, but not necessarily longer than the length scales
associated with the inhomogeneities of the system~\cite{burridge__1967, olami_self-organized_1992}. For long range stress transfer without inhomogeneities, as well as randomly distributed inhomogeneities~\cite{serino_gutenberg-richter_2010} such models have been found to produce
scaling similar to Gutenberg-Richter scaling found in real earthquake systems~\cite{gutenberg__1956}.  It has been shown that the scaling found in such models is due to a spinodal in the limit of long range stress transfer~\cite{rundle_scaling_1993, klein_geocomplexity}.  

In the earthquake lattice models we use in this work we introduce inhomogeneities in the way that stress is dissipated. 
Stress is dissipated both at the lattice site of failure (site dissipation) and at neighboring sites which are damaged (damage dissipation).  Spatial inhomogeneities are incorporated by varying this stress dissipation throughout the system in different spatial arrangements. We find that the scaling for damaged systems depends not only on the amount
of damage, but also on the spatial distribution of that damage as well as the relation of the spatial damage or dissipation to the stress transfer range.  Studying the effects of various spatial arrangements of site dissipation provides insights into how to construct a realistic model of an earthquake fault which is consistent with Gutenberg-Richter scaling.

\section{Model}

We use a two-dimensional cellular
automaton model of an earthquake fault which is a variant of the RJB model~\cite{rundle_1977, rundle_origin_1991} and closely resembles the OFC model~\cite{olami_self-organized_1992}. We begin with a two-dimensional lattice, where each site is
either dead (damaged) or alive (active). Each live site $i$ contains
an internal stress variable, $\sigma_{i}(t)$, which is a function
of time. All stress variables are initially below a given threshold stress,
$\sigma^{t}$ and greater than or equal to a residual stress $\sigma^{r}$
(both of which we assume to be spatially homogeneous.)
Sites transfer stress to  $z$  neighbors.
Neighbors are defined as all sites within the transfer range, $R.$ 
Initially we randomly distribute stress to each site so that $\sigma^{r}<\sigma_{i}<\sigma^{t}$.
We then increase the stress on all sites equally until one site reaches $\sigma^{t}.$
At this point, the site at the threshold stress fails. When a site fails, some fraction of that
site's stress, given by $\alpha_{i}(\sigma^{t}-\sigma^{r}\mp\eta),$ is
dissipated from the system, where $\alpha_{i}$ is a parameter that characterizes the fraction of stress dissipated from site $i$, and $\eta$ is a random flatly distributed noise. The stress of the site is lowered
to $\sigma^{r}\pm\eta$ and the remaining stress is distributed equally
to the site's $z$ neighbors.

To model more realistic faults, we use systems which are \emph{damaged}, meaning they have both alive sites, which obey the rules outlined above, and dead sites which do not hold any stress.  Following Serino, et al~\cite{serino_cellular_2010}, in addition to the stress dissipation regulated by the site dissipation parameter, $\alpha_{i}$, we specify that any stress which is passed to a neighboring dead
site also gets dissipated from the system. We can therefore regulate the spatial distribution of stress dissipation from the system with the distribution of the $\alpha_{i}$ and the placement of dead sites on the lattice. After the initial site
failure, all live neighbors are then checked to see if their stress
has risen above $\sigma^{t}$. If it has, this site goes through the
same failure procedure outlined above until all sites have stress
below $\sigma^{t}$. The size of the avalanche is the number
of failures that stem from the single initiating site. We refer to this whole avalanche
process as a plate update.

Because stress is dissipated from the system both at the site of failure (as
regulated by $\alpha_{i}$) and
through dead sites which may be placed inhomogeneously  throughout the system, we may think of each site $i$ as having a parameter which incorporates both types of dissipation, $\gamma_{i}=1-\phi_{i}(1-\alpha_{i})$,
where $\phi_{i}$ is the fraction of live neighbors of site $i$.   The mean value $\overline{\gamma}=\sum_{i}\gamma_{i}/N_{a}, $where $N_{a}$ is the number of live sites, is the average fraction of excess stress dissipated from the system per failed site. 

We will want to compare the scaling in these systems with the scaling in systems where the damage distribution is uniform. 
It has been found ~\cite{klein_structure_2007} for these OFC type models with no spatial inhomogeneities (homogeneous damage and constant $\alpha_{i}$) that in the mean field limit the number of avalanche events of size s obeys the scaling form 
\begin{equation}
n(s)\sim e^{-\Delta h\,s}/s^{-\tau}.\label{eqn:ns}
\end{equation}
The quantity $\Delta h$, which is a function of the fraction of dead sites, is a measure of the distance from the spinodal and $\tau=3/2$. (Note that $n(s)$ is the number of events of size $s$, which is the non-cumulative distribution, rather than the number of events of size $s$ or smaller, which is the cumulative distribution often discussed in relation to the Gutenberg-Richter law.)  We know from Ref.~\cite{serino_cellular_2010} that long range damaged systems with a fraction $\phi$ of live sites and constant site dissipation parameter $\alpha_{i}$ are equivalent to undamaged systems with site dissipation parameter $\alpha' = 1-\phi(1-\alpha).$ These systems approach the spinodal ($\Delta h \rightarrow 0$) as the stress dissipation from the system vanishes: $\phi \rightarrow 1$ and $\alpha\rightarrow 0.$  Physically, stress dissipation from the lattice system suppresses large avalanche events.

\section{Qualitative behavior of scaling}

\begin{figure}[htp]
	\begin{center}
    		\subfigure[\rd]{\label{fig:drd}\includegraphics[scale=0.35]{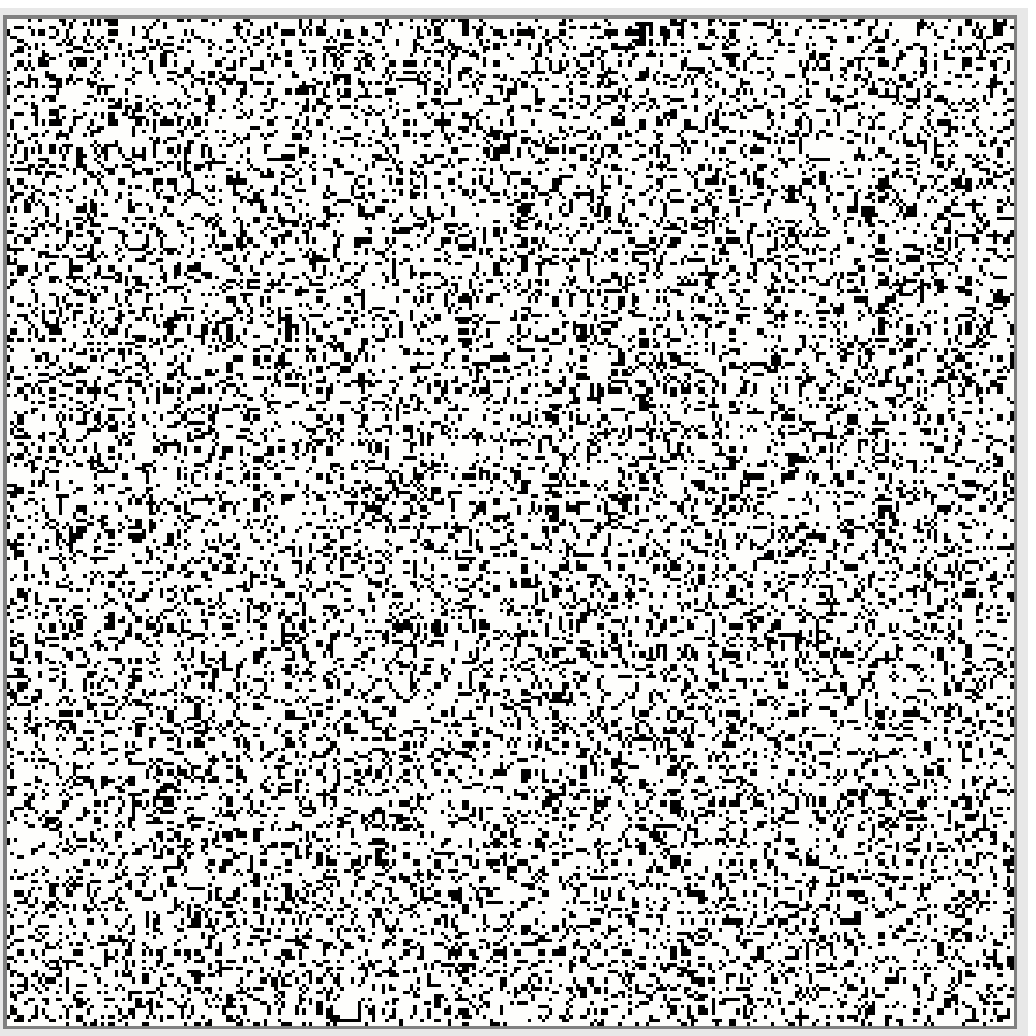}}
    		\subfigure[\rc]{\label{fig:dcr}\includegraphics[scale=0.35]{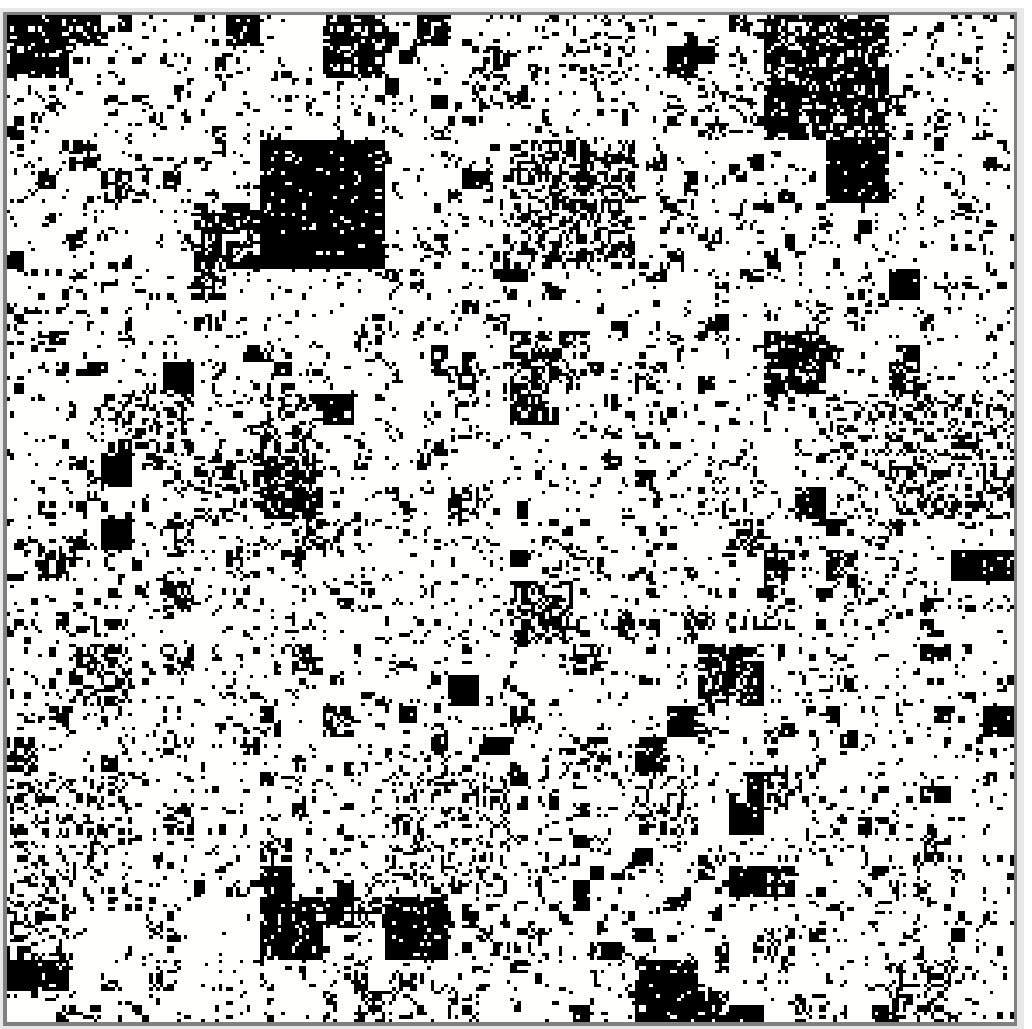}} \\
    		\subfigure[\cb]{\label{fig:dcb}\includegraphics[scale=0.35]{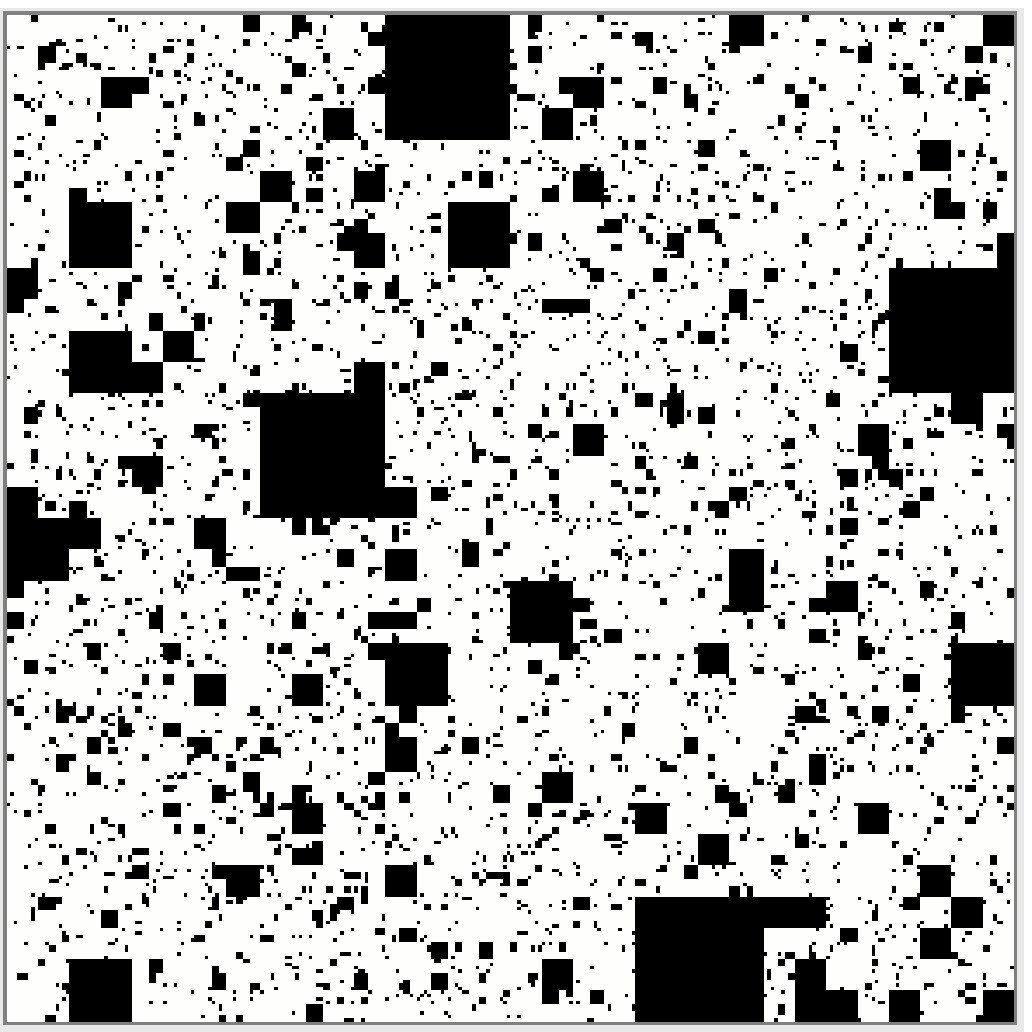}}
    		\subfigure[\bd]{\label{fig:d16}\includegraphics[scale=0.35]{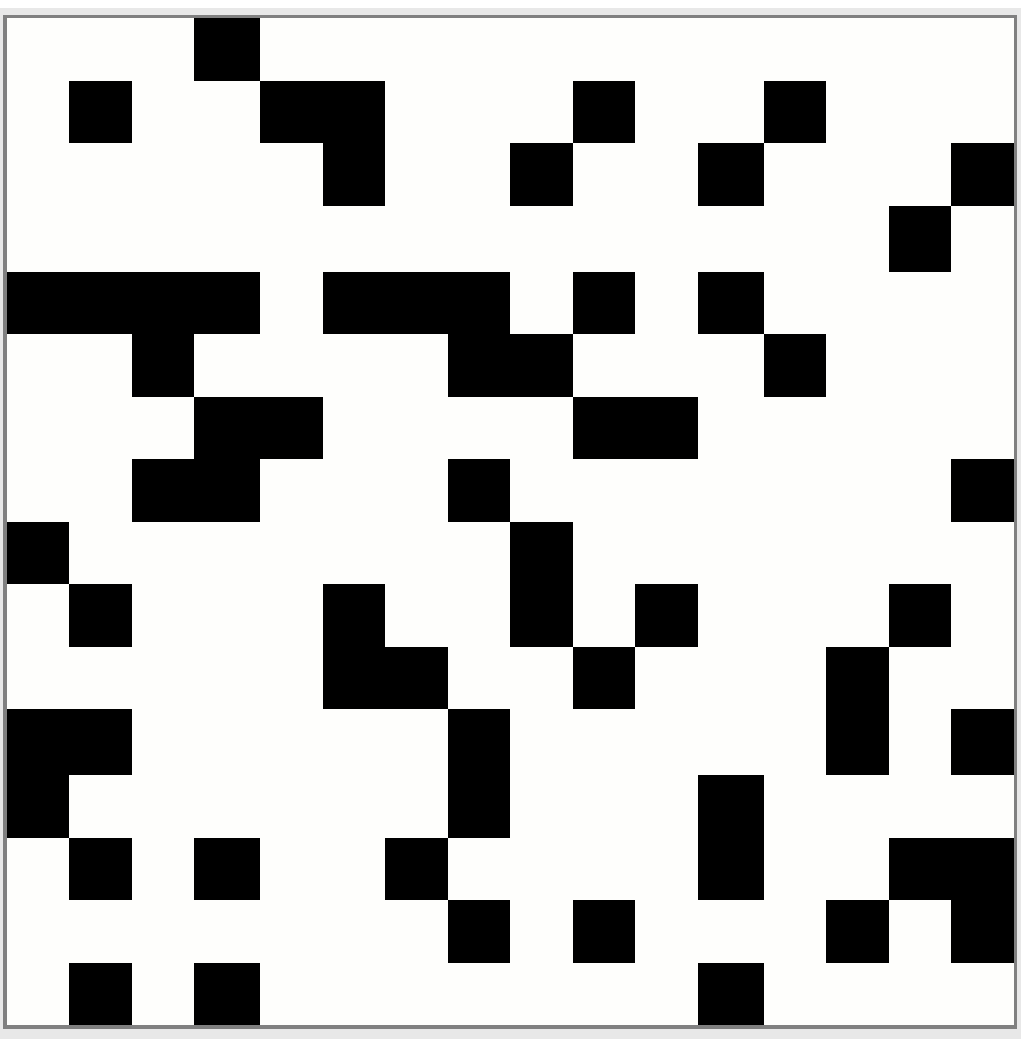}}
  	\end{center}
 	\caption{Various configurations of 25\% dead sites (in black) for a lattice with linear size $L=256$.  Lattices contain (a) dead sites distributed randomly, (b) blocks of various sizes, where each block has $p$ randomly distributed dead sites with $p$ varying for each block, (c) dead blocks of various sizes,  (d) dead blocks of a single size.\label{fig:damageTypes}}
\end{figure}

First we study the case with constant $\alpha_{i}=\alpha$ and damage distributed inhomogeneously throughout the system. 
In Fig.~\ref{fig:damageTypes} we show two dimensional lattices of linear size $L=256$ and 25\% of the sites dead. The lattices have various distributions of the dead sites. 
Figure~\ref{fig:drd} has the dead sites randomly distributed throughout the system.  In the long range limit, this corresponds to homogenous damage studied in Ref.~\cite{serino_cellular_2010}.  Figures~\ref{fig:dcr}-~\ref{fig:d16} incorporate some clustering of dead sites.  Figure~\ref{fig:dcr} has blocks of randomly distributed dead sites with cascading length scales where the linear block sizes range from $1$ to $L/8.$  Each block has a fraction $p$ of randomly distributed dead sites, where $p$ varies from block to block. The values of $p$ are selected from a random Gaussian distribution.  Figure~\ref{fig:dcb} has dead blocks with cascading length scales where the linear block sizes also range from $1$ to $L/8.$  Figure~\ref{fig:d16} has randomly distributed dead blocks with blocks of linear size of $L/16$ only.  To characterize each configuration in Fig.~\ref{fig:damageTypes}, we calculate $\overline{\gamma},$ and the variance of $\gamma_{i}$ for an interaction range $R=16$ and $\alpha_{i}=0\,\forall i.$  The results are summarized in Table~\ref{tab:summary_table}.

\begin{table}
  \begin{tabular}{ | c | c | c| }
    \hline
    Damage Distribution & $\overline{\gamma}$ & Variance\\ \hline
     \rd& $0.2510$ & $2.1\times 10^{-4}$\\ \hline
     \rc  & $0.2293$ & $6.9\times 10^{-3}$\\   \hline
     \cb & $0.2092$ & $8.9\times 10^{-3}$\\ \hline
     \bd & $0.1803$ & $2.0\times 10^{-2}$\\ 
    \hline
  \end{tabular}
  \caption{Averages and variances of $\gamma_{i}$ for the distributions of dead sites shown in Fig.~\ref{fig:damageTypes}. The total number of dead sites is equal to 25\% of the lattice for all distributions.}
  \label{tab:summary_table}
\end{table}

\begin{figure}
\includegraphics[scale=0.7]{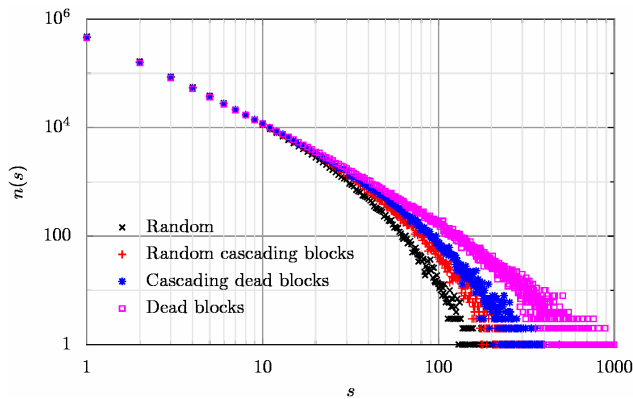}\caption{Numerical distribution of avalanche events of size $s$ for various spatial distributions of dead sites.  Data corresponds to lattices in Fig.~\ref{fig:damageTypes} with interaction range $R=16. $\label{fig:damageDistPlot}}
\end{figure}

Figure~\ref{fig:damageDistPlot} shows $n(s),$ the numerical distribution of avalanche events of size $s,$  corresponding to the various distributions of damage in Fig.~\ref{fig:damageTypes}.  We find that the scaling behavior of systems with damage depends not only on the total amount
of damage to the system but also on the spatial distribution of damage.  In particular, large events are suppressed more for lattices with damaged sites distributed more homogeneously.  Because these lattices are of equal size, have the same number of damaged sites and the same stress transfer range($R=16$)
 the differences in the large event behavior are not due to the finite size of the lattice or the finite number of active sites in the lattice. Furthermore, the results of Sec.~\ref{sec:spatialDists} indicate that the effect is due to the spatial distribution of $\gamma_i$s and does not even require that the lattice be damaged.
The calculated quantities in Table~\ref{tab:summary_table} would appear to indicate that the large event suppression is correlated with both higher values of the average dissipation parameter $\overline{\gamma}$ and lower values of the variance of $\overline{\gamma}.$

In order to better understand these results, we now study the effect of the interaction range relative to the length scales of inhomogeneities and the effect of clustering of dead sites.

\subsection{Length Scales}

For any given distribution of damage, the system will act as if the damage is homogeneous if the stress transfer range is long enough compared to the length scales of damage of the lattice.  To illustrate the importance of relative length scales, we consider the case of a single length scale associated with damaged areas.  We place blocks of damaged sites of size, $b$, randomly in the system which has constant $\alpha_{i}=\alpha.$ See, for example, Fig.~\ref{fig:d16}.  As we vary the ratio $R/b$, the measured value of $\overline{\gamma}$ varies from $\overline{\gamma}=\alpha$ for $R/b\ll 1$ to $\overline{\gamma}=1-\phi(1-\alpha)$ for $R/b\gg 1$.  In the former case, the live domains of the system appear nearly homogeneous with $\phi_{i} = 1$ except near the boundaries of dead blocks.  The latter case is the limit of homogeneously distributed damage.  In both limiting cases, the variance of $\gamma_{i}$ is small and the scaling is equivalent to the scaling for an undamaged system with $\alpha'=\overline{\gamma}$.  

\begin{figure}
\includegraphics[scale=0.7]{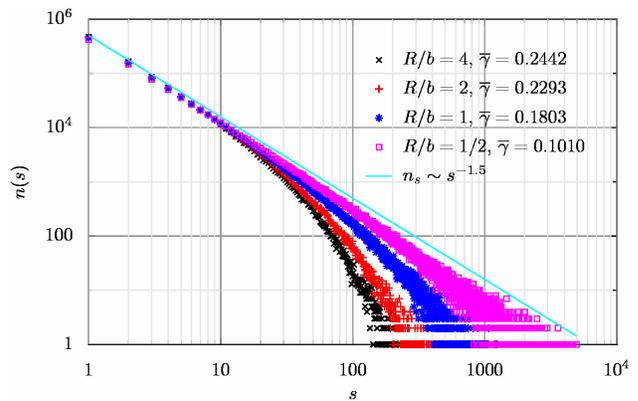}\caption{Numerical distribution of avalanche events of size $s$ for blocks of dead sites of linear size b. (Figure~\ref{fig:d16} corresponds to $R/b=1.$)   The size of the system is $L=256$ and the interaction range is $R=16.$  The line is drawn to show that the data is approaching a power law with exponent $-3/2.$  \label{fig:blockCompare}}
\end{figure}

In Fig.~\ref{fig:blockCompare}, we compare systems with randomly distributed
dead blocks of various length scales, $R=16,$ $\alpha=0,$ and 25\% total damage.  As $R/b$ gets small, the values of $\overline{\gamma}$ also get small. The distribution $n_{s}$ of the corresponding data approaches a  power law with the exponent $-3/2,$ which is the form of the distribution of a system at the spinodal.

\subsection{Spatial Distributions of Dissipation\label{sec:spatialDists}}

The spatial distribution of damaged sites determines the spatial distribution of $\gamma_{i}$ values.  A more direct way to control the numerical and spatial distributions of $\gamma_{i}$ is to use undamaged systems and vary the values of $\alpha_{i}$.  In this way, we can isolate the effects of spatial redistribution of $\gamma_{i}$ values while holding the numerical distributions of $\gamma_{i}$ constant.

\begin{figure}[htp]
	\begin{center}
    		\subfigure{\label{fig:aDist}\includegraphics[scale=0.7]{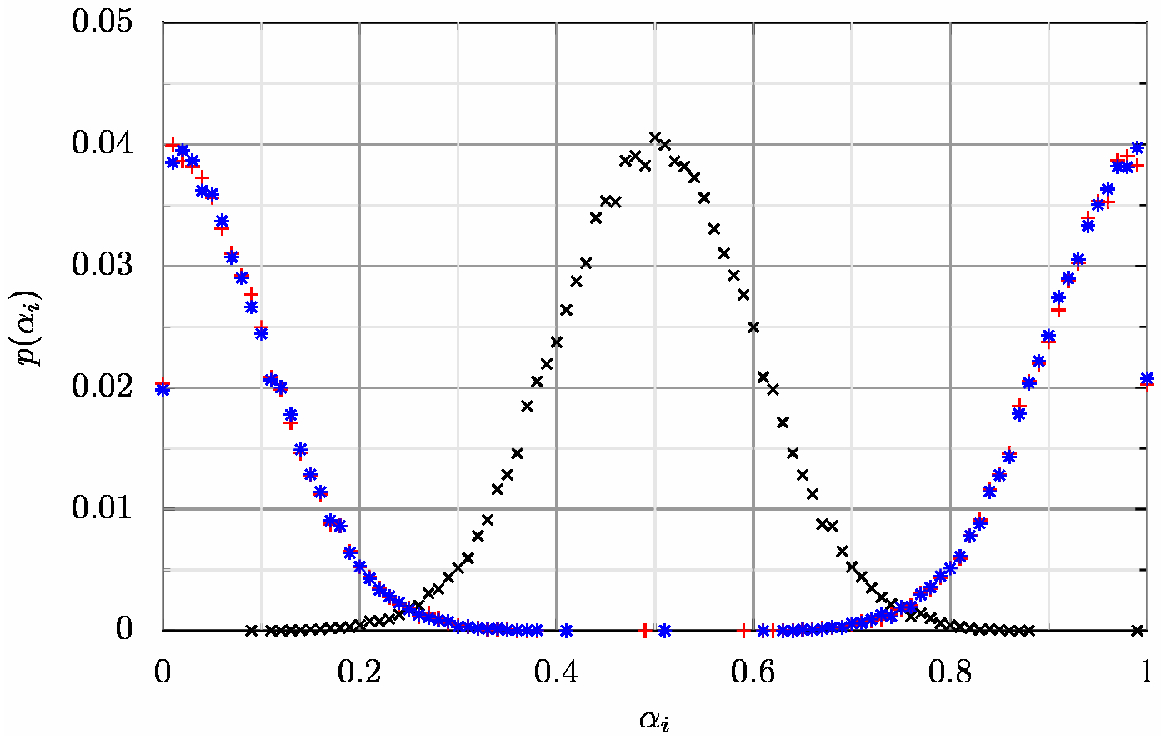}}\\
    		\subfigure{\label{fig:spatialDist}\includegraphics[scale=0.7]{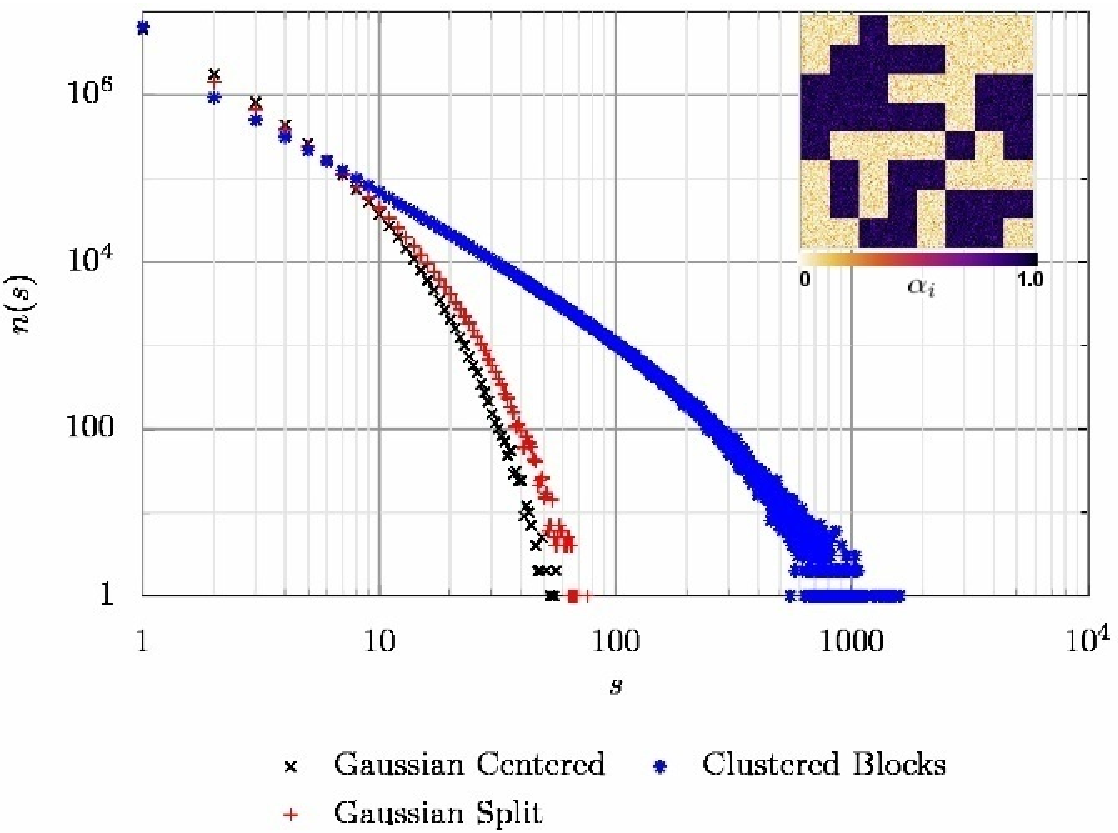}}
  	\end{center}
 	\caption{Comparison of three lattice systems with no damage and two different distributions of $p(\alpha_{i})$ shown in the top figure. The lattices labeled by ``Gaussian Centered'' and ``Gaussian Split'' are distributed uniformly in space, while the spatial distribution of ``Clustered Blocks'' is shown in the inset.  The bottom plot shows the numerical distribution of avalanche events of size $s$.}
	\label{fig:alphaDist}
\end{figure}

We present data in Fig.~\ref{fig:alphaDist} for three systems with site dissipation only; that is, they have no damage and $\gamma_{i}=\alpha_{i}$ for each system.  We see that $\overline{\gamma}=0.5$ for all three systems from the numerical distributions of $\alpha_{i}$ values, $p(\alpha_{i})$.  

The two systems labeled ``Gaussian Split'' and ``Gaussian Centered'' both have a uniform spatial distribution of $\alpha_{i}$ values.  However, as shown in Fig.~\ref{fig:alphaDist}, the values of $\alpha_{i}$ for the ``Gaussian Centered'' system have a Gaussian distribution centered about $\alpha_{i}=0.5,$ while the values of $\alpha_{i}$ for the ``Gaussian Split'' system have partial Gaussian distributions and are clustered near the values of $\alpha_{i}=0$ and $\alpha_{i}=1$.  Thus, the variance of  $\alpha_{i}$ values for the ``Gaussian Centered'' system is less than the variance for the ``Gaussian Split'' system.  We see from the numerical distribution of avalanche events, $n(s),$ in Fig.~\ref{fig:alphaDist} that the ``Gaussian Split'' system has slightly larger events than the ``Gaussian Centered'' system, consistent with the observations above that large event suppression correlates with low variances of $\alpha_{i}.$  

However, we see by studying the ``Clustered Blocks'' system that spatial distributions of $\alpha_{i}$ have a robust effect on scaling, even when the variances of $\alpha_{i}$ are the same. The ``Gaussian Split'' and ``Clustered Blocks'' systems have nearly the same numerical distributions of $\alpha_{i}$ (Fig.~\ref{fig:alphaDist}), and therefore have the same value of the variance of $\alpha_{i}.$  The spatial distributions of these two cases, however, are different: the ``Gaussian Split'' system has a uniform spatial distribution of $\alpha_{i}$ values, while the ``Clustered Blocks'' system has high  (and low) $\alpha_{i}$ values clustered together into blocks as shown in the inset of Fig.~\ref{fig:alphaDist}.  Despite having equal values of $\overline{\alpha_{i}}$ and equal variances of $\alpha_{i}$ values, the ``Clustered Blocks'' system experiences much larger events (by an order of magnitude).  

Evidently, the larger events depend crucially on the spatial clustering of low dissipation sites.  This is because failing sites with low values of  $\gamma_i$ pass along a high percentage of excess stress, encouraging the failure of neighboring sites. Thus, a large earthquake event is more likely to occur if the initial site of failure is well connected to a large number of sites with low dissipation parameters.  In our system, connectedness is determined by spatial locality, so we require large clumps of sites with low values of $\gamma_{i}$ in order to allow for the occasional large earthquake event.

\section{Gutenberg-Richter Scaling}

The Gutenberg-Richter scaling law states that the cumulative distribution of earthquake sizes is exponential in the magnitude~\cite{gutenberg__1956}.  In terms of the seismic moment, which has succeeded the Richter magnitude as the appropriate measure for earthquake sizes, the law may be reframed to state that the cumulative distribution of earthquake sizes, $N_{M},$ is a power law in the seismic moment, $M$~\cite{serino_gutenberg-richter_2010}.  
\begin{equation}
N_M\sim M^{-\beta}\,,\hspace{1mm}\text{with}\hspace{2mm}\beta\equiv\frac{2b}{3}\,,
\label{eqn:gr}
\end{equation}
and $b$ is the so-called $b$ value of the Gutenberg-Richter law which has been measured for many real earthquake systems.  The seismic moment $M$ is proportional to the size of the earthquake in this model~\cite{rundle_geocomplexity_2000}. Therefore, the relation appropriate for the systems considered in this work is the cumulative distribution of earthquake size:
\begin{equation}
N_s\sim s^{-\beta},
\label{eqn:sizeScale}
\end{equation}
or the corresponding non-cumulative distribution
\begin{equation}
n_s\sim s^{-\tilde{\tau}}\,,\hspace{1mm}\text{with}\hspace{2mm}\tilde{\tau}=\beta+1.
\label{eqn:sizeScale}
\end{equation}

Serino et al ~\cite{serino_gutenberg-richter_2010} construct a model for an earthquake fault system consisting of an aggregate of lattice models, where each lattice has a fraction $q$ of homogeneously distributed dead sites and $q$ varies from $0$ to $1.$ The weighting factor $D_{q}$ gives the fraction of lattices with damage $q$.  Considering the weighting factor to be constant with all values of $q$ contributing equally to the fault system, they find a value of $\tilde{\tau}=2.$ They also consider a power law distribution of $D_{q}$ and fit the exponent to correspond to Gutenburg-Richter $b$ values found in real earthquake systems.  

There are two important differences between the model considered by Serino et al and our work: In the model treated by Serino et al
\begin{enumerate}
\item The damage is distributed homogeneously.
\item The individual lattices with homogeneous damage $q$ are non-interacting.
\end{enumerate}
We investigate both the effect of the spatial arrangement of the damage and its relation to the stress transfer range as well as the effect of stress transfer between regions with different levels of damage.

\begin{figure}[htp]
	\begin{center}
    		\subfigure[]{\label{fig:many}\includegraphics[scale=0.7]{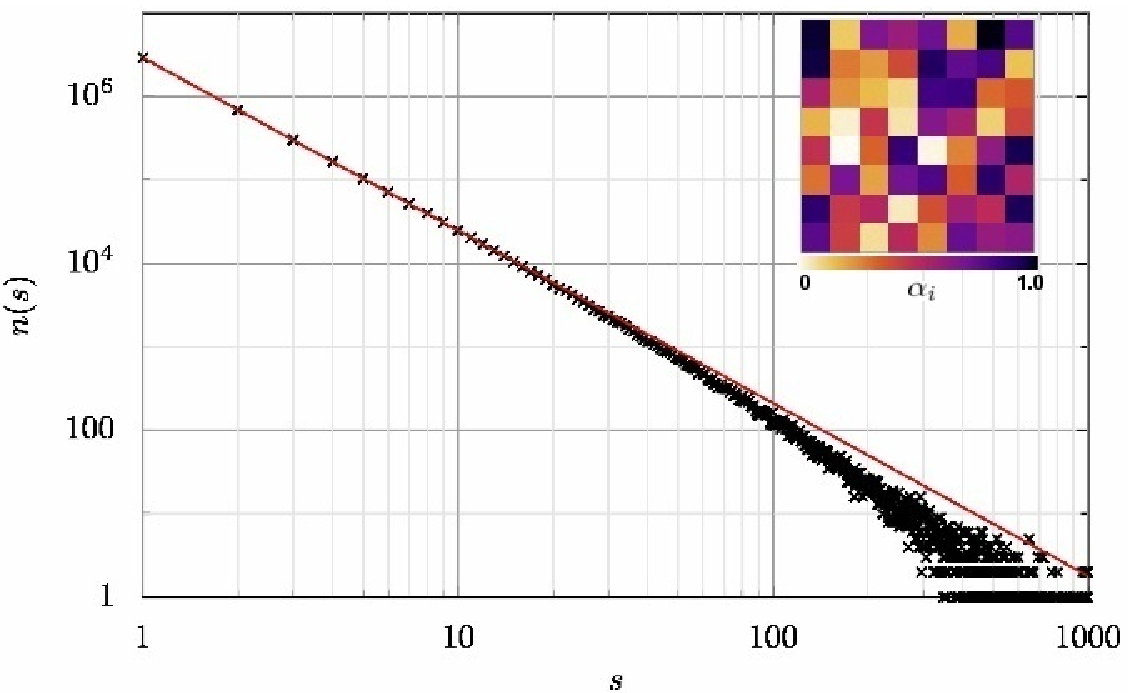}}\\
    		\subfigure[]{\label{fig:eights}\includegraphics[scale=0.7]{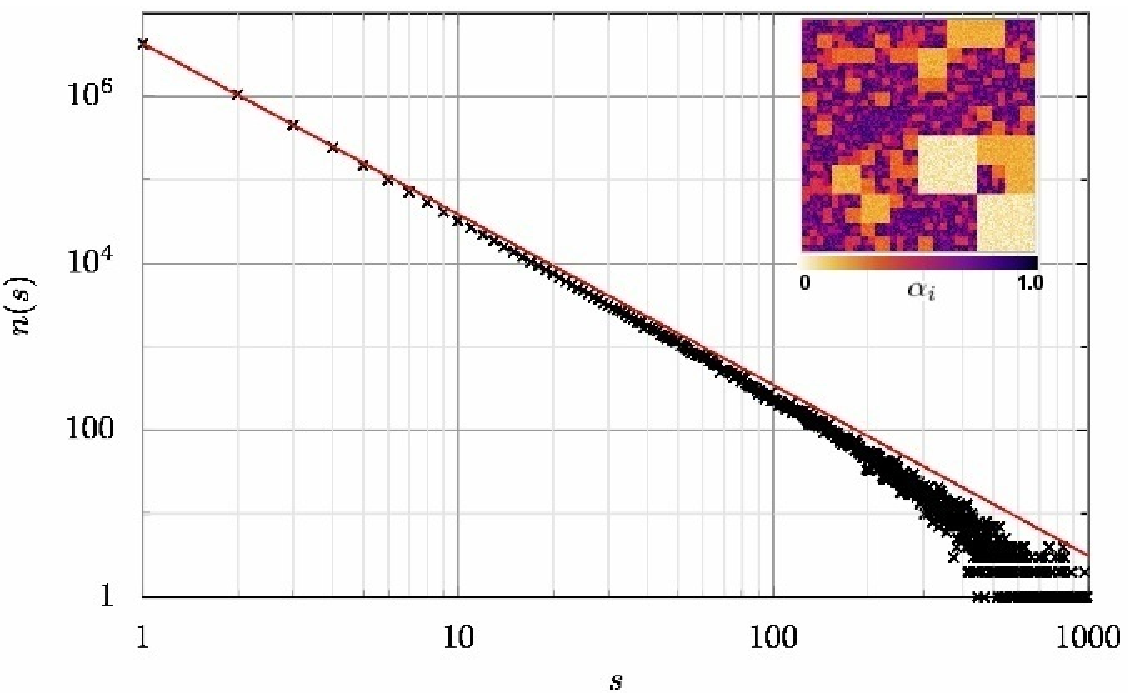}}
  	\end{center}
 	\caption{Numerical distribution for avalanche events of size $s$ for systems with uniform numerical distributions for $\alpha_{i}$, but non-uniform spatial distributions of $\alpha_{i}$ which are shown in the insets.  Slopes of best fit lines in red are a) $\tilde{\tau}\simeq2.07$ and b) $\tilde{\tau}\simeq2.04.$}
	\label{fig:GRlattices}
\end{figure}

 We construct two lattice systems with a uniform numerical distribution of $\gamma_{i}$ which have scaling consistent with Serino et. al.'s systems with constant distribution $D_{q}.$
The first model essentially pieces together many homogeneous lattice systems: The numerical distribution of $\alpha_{i}$ values is uniform between $0$ and $1$, but spatially arranged into $N_{B}$ blocks of linear size $B$ (see Fig.~\ref{fig:many} inset), where each block contains a random distribution of $\alpha_{i}$ values within an interval of size $1/N_{B}.$ There are no dead sites, so that $\alpha_{i}=\gamma_{i}.$  The effects of the boundaries between the blocks should be negligible if $B\ll R.$  In Fig.~\ref{fig:many}, we present data from a system with $L=512,$ $R=16,$ and $B=64.$  The straight line shows the best fit to a power law with exponent $\tilde{\tau}\simeq2.07,$ which is consistent with the results for the aggregate lattice system of Serino et al with $D_{q}=1$.

We find that the size of the blocks, $B$, need not be the same for different values of $\alpha_{i}$.  It is important that the boxes with lower values of $\overline{\gamma}$ be large enough to accommodate large avalanche events, but blocks with large $\alpha_{i}$ may be small because they are more likely to seed small avalanches.  With this in mind we construct a lattice system with cascading length scales of blocks where the largest blocks have the lowest $\alpha_{i}$ values and decreasing sized blocks have increasing values of  $\alpha_{i}.$  The scaling results are shown in Fig.~\ref{fig:eights}, with a best fit power law with exponent $\tilde{\tau}\simeq2.04.$

\section{Conclusions}

We have studied both damage and site dissipation to inform the development of models of realistic earthquake faults with inhomogeneous stress dissipation.  Spatially rearranging dead sites on a given lattice affects the numerical distributions of the effective stress dissipation parameters and the scaling behavior of large avalanche events, depending on the homogeneity of the damage and the length scales associated with the clustered dead sites.  However, by studying site dissipation we find that spatial distributions of dissipation parameters crucially affect scaling behavior even when the numerical distributions of dissipation parameters are the same.

Sites with lower stress dissipation, even if only partially distributed throughout the lattice but clumped together, allow for larger avalanche events.  We have found models for earthquake fault systems which have avalanche event size scaling which is  consistent with the new paradigm for Gutenberg-Richter scaling proposed by Serino et al.  The models studied here
go beyond those previously proposed by incorporating inhomogeneities into the lattice and allowing areas with different characteristic dissipation rates to interact.

\begin{acknowledgments}
This work was funded by the DOE through grant DE-FG02-95ER14498 and the NSERC and Aon Benfield/ICLR Industrial Research Chair in Earthquake Hazard Assessment.

\bibliography{scalingPaperBib}{}

\bibliographystyle{apssamp.bst}

\end{acknowledgments}

\end{document}